Short Paper

# RUI: A Web-based Road Updates Information System using Google Maps API

Benzar Glen S. Grepon
Northern Bukidnon State College, Philippines
ben.it2c@gmail.com
(corresponding author)

JC P. Margallo
Northern Bukidnon State College, Philippines

Jonathan B. Maserin
Northern Bukidnon State College, Philippines

Rio Al-Di A. Dompol
Northern Bukidnon State College, Philippines

## Abstract

*Purpose* – Knowing the current situation on every road in an area is still difficult to anticipate. Commuters, riders, and drivers are still dependent on road situations from a local news agency to be well informed and be updated on possible road updates such as vehicular accidents, government road and bridge projects/construction, and other related road obstructions. To give solutions regarding road updates, a web-based roads update information system has been developed that uses Google Maps API allowing people to view and be notified of the real-time updates of the road situation of a specific area.

*Method* - This paper discusses the main system functionalities, including sub-systems and modules of the system, the research approach and methodology, which is the Agile Model, and its impact on disseminating road information and its status.

*Results* – The project has been evaluated using ISO 25010, a quality evaluation system for product/software measuring its functionality, efficiency, reliability, usability, security, maintainability, and portability. Based on the evaluation result, the project has been rated 4.21, signifying an excellent performance based on qualitative description through a Likert scale descriptive interpretation. The project has been running and hosted on the world wide web and is expected to expand its coverage area from its origin country to the rest of the world.

*Conclusion* – Based on the initial findings of the study, the respondents agreed that the developed web system was functional and a massive help to commuters, riders, and people who travel a lot. The system's overall effectiveness and performance were excellent based on the criteria set by ISO/IEC 25010.

*Recommendations* – It is recommended for future development to expand the coverage of the road updates, if possible, including the entire Philippine archipelago for long-drive commuters and drivers to be more updated in terms of road updates. Also, include the use of mobile applications for more user-friendly design and interactions.

*Research Implications* – The study allows viewing and reporting road updates with attachments as proof, informing commuters about what to expect when they travel. The system has been implemented in the municipality of Bukidnon and has been updated now and then for continuous quality improvement.

*Practical Implications* – The main benefit that the Road Updates Information System can do is to inform people who are constantly traveling to be updated on the current situation of a specific road by accessing the system and checking the map on areas to travel and see for themselves the photos and description of a place where road construction, accidents and other causes that delay the traveling time of passengers and travelers. Aside from being informed of the road updates, citizens can also participate by uploading photos and descriptions of what is still going on the roads to the system and be viewed by people who try to access the web portal.

*Keywords* – Road Updates, Information System, Google Maps, Web App, Web Portal, Philippines

## INTRODUCTION

With growing municipality dwellings, expanding roots of human settlements, and the complex urbanization happening all over, the way we find our way through routes and off-the-routes is getting tougher. This is why navigation apps are growing in prominence and scope to help us better read, predict, and pierce through streets and directions (Yamsaengsung and Papasratorn, 2018). People use navigating apps as their primary information access (Wörndl and Herzog, 2020) to familiar and unfamiliar destinations or personal routes (Löchtefeld, 2019).

Knowing the current situation of a particular place, especially road conditions, and updates, is difficult to monitor with the existing navigation apps. The problem began when smartphone apps like Waze, Apple Maps, and Google Maps came into widespread use, offering drivers real-time routing around traffic tie-ups; an estimated 1 billion drivers worldwide use such apps. Now that online navigation apps are in charge, they are causing more problems than they solve (Macfarlane, 2019). In the case of the Philippines and experienced in other countries, there are a lot of on-going situations that cause delays (e.g., landslides, on-going road repairs, accidents, and others) to commuters and travelers' day by day. This situation has inspired the proponents to design and develop a Road Updates Information System. It's still challenging to predict the state of every road in each area. To stay informed about potential road updates such as traffic accidents, governmental road and bridge projects/construction, and other associated road impediments, commuters, riders, and drivers continue to rely on the road status provided by a local news organization. Traveling without the idea of what's going on ahead will make the traveling riskier; for example, planning to go somewhere where there is on-going rescue operation because of the landslide currently happening or delay because there are roads and bridges needed to be repaired, these incidences will undoubtedly delay the traveling time and expectations of a particular traveler or commuter which may affect their transactions, vacations or any essential dealings.

Fortunately, the proposed system has the capabilities to help riders, drivers, commuters, or even ordinary people to be able to navigate the map through a web application and be updated on what's happening, aside from being informed those who were in the actual site where accidents happened, landslide and other factors which delay the passing by of vehicles in certain places can participate by updating the map and upload pictures and information to update the maps with the present situation. This web app uses Google Maps API to locate the exact location of a condition that causes the delay and inform all web app users of the situation, then giving them the decision whether to continue to travel or reroute to other alternative routes to avoid being congested where any incidences happen.

## LITERATURE REVIEW

Navigation apps have been very beneficial to most, given that it helps us navigate places to which we are familiar or unfamiliar. Currently, no direct study is very similar to the study being presented. A couple of the studies are listed below that use the same technology and methods the proponent could relate to.

The study of (Yang and Hsu, 2016) uses google maps API, which were used by this study to locate specific area point. Yet, it uses image recognition technology as a tourism information provider and tourist route planner. Another similar is the Study of (Luthfi, et al, 2019) Wherein they made use of iOT Technology, allowing their objects to connect to the internet for them to track and monitor, then used Google Maps API to carry out their project through direct web access or mobile phone. A study by (Khoo and Ong, 2013) presents the development of a holistic structural equation model that can consider all potential attributes affecting perceived quality in a single model for a traffic information system which is one of the factors with relates to the delays in the roads for travelers. Another study wherein it uses a GPS navigation system based on google maps API has been studied and then implemented by (Li & Zhijian, 2010), which uses Google Maps to navigate using mobile devices.

There is a study that allows the participation of ordinary people concerning incident mapping is the study of (Balahadia, et al., 2015), where citizens are allowed to point areas in the community through a digital map that requires attention from the barangay officials; they used social media integrations for the community to receive notifications for every update done by the officials to the website. Another study where citizens' participation in mapping has been integrated is in the study of (McCall, 2021), which distinguishes facts and other media uploaded by people to avoid fake news and the like. The use of social media as an identifier for natural disasters has been a tool used by citizens to report incidences and inform disaster sites and safe points for casualties evacuation in real time with GIS aid a study of (Slamet, et al, 2018) which implements Secure Place Locator (SPL) which can map position, quantity, density, and incident in the disaster site so that Disaster Management can be performed quickly and accurately.

Regarding road safety, the study of (Dadvar, et al, 2020) has a database that maintains motor vehicle crash data, roadway inventory, and traffic volume data for several US States, one of the target functionalities of the Road Updates Information system. A study of (Novikov, et al, 2021) uses Geoinformation System (GIS) looking to the various factors affecting road safety, accident likelihood, and severity of their consequences depending on the hazardous areas location, which is the main target of the system to identify potential factors which cause a delay in the road and allow participation from the public to notify travelers with regards to the current condition of the road where incidences happen. Recent studies have revealed that navigation systems frequently guide users down hazardous roads under the false assumption that all roads are equally safe. A study of (Xu, et al, 2022) developed a machine learning-based road safety classifier that predicts the safety level for road segments using a diverse feature set constructed only from publicly available geographic data. Two (2) mobile communications services technologies are used in the study of (Thalluri, et al, 2022) GPS for vehicle tracking and GSM to transmit and update the vehicle's location into the database. In addition, they use API Key to display the vehicle on a map on the web application, and it also makes use of an ultrasonic sensor to warn the driver of the presence of any adjacent vehicles and help him avoid collisions and potential accidents while navigating the road safely. Lastly, in terms of the use of localized qualitative traffic information with an ontological model for traffic information, the study of (Amado and Dela Cruz, 2022) utilized an integrated sensor system to gather data on the temperature, relative humidity, rainfall intensity, and flood level of the deployment site. A Haar cascade deep learning algorithm to extract traffic volume information and accident information from video feeds.

## METHODOLOGY

### *Software Development*

Over the years, SDLC has remained the reliable approach to software development (Khan et al., 2020). The Agile technique, as indicated in Figure 1, is ideally suited for speedy and effective software development due to its adaptive nature, early delivery, and flexible lifecycle (Srivastava, Bhardwaj & Saraswat, 2017).

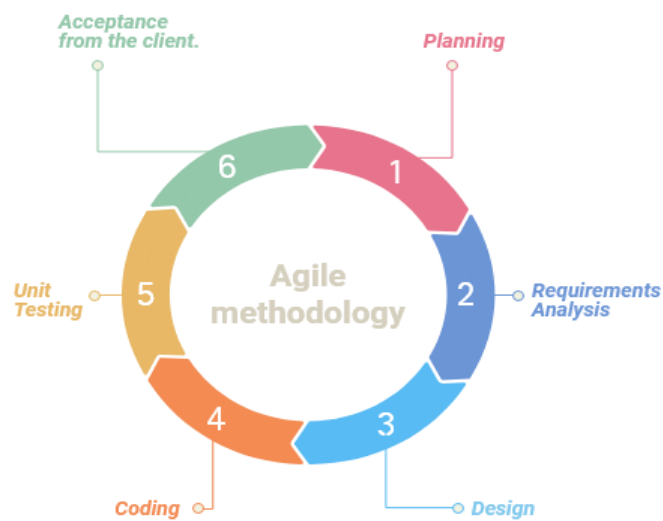

*Figure 1*. Agile Methodology in System Development

## *Planning*

The researchers obtained data in a Municipality of Bukidnon. They examined the typical issues they ran into. According to the report, many people have accidents because they lack information; thus, they require information to warn them of accident-prone areas, roadblocks, and on-going construction.

## *Requirements Analysis*

To analyze the requirements using some of the many system flow diagrams and flow charts available. Data flow diagrams are used to represent the flow and to help advocates better understand how to put systems into place and ensure that data is flowing accurately. Additionally, flowcharts are used to demonstrate the system's process, which aids the proponents in making clear how things will operate. The process that needs to be carried out in an operation is depicted using a flowchart as a series of steps.

## *Design*

In this phase, the researcher plans to make a good design for the project. The researcher surveyed random riders and commuters to gather data information to collect the essential needs of this system. The proponents came up with a functional design of the Network Infrastructure, a framework for the specification of a network's physical components and their functional organization and configuration; the System Architecture, which is the conceptual model that defines the structure, behavior, and more views of a system, the Database Model that determines the logical structure of a database and fundamentally determines in which manner data can be stored.

## *Coding*

In this phase, the development and writing of the actual code, the researcher uses WordPress to develop web apps; WordPress is the most popular open-source Content Management System (CMS), used by approximately 75 million websites. The researcher studied the hypertext processor (PHP) as the programming language used for web development. The researcher also uses CSS cascading style sheets for creative design to look better. Lastly is JavaScript (JS) for animation and interaction.

## *User Testing*

The researcher must repeat this procedure until no errors are identified in the functionality of the software development at this stage, where it involves the designing and coding processes.

## Acceptance for the Client

The researchers presented the output web application to propose a random user for acceptance through mobile browser demonstration with System Evaluation Scale (SUS). A mobile browser demonstration will be used as the introductory and brief flow of the application.

## System Testing and Evaluation

During the evaluation of the regular users, it gives the feedback score about the system of the final output. ISO/IEC 25010 is used for the application evaluation assigned to the participants because this ISO Standard provides a uniform vocabulary for defining, assessing, and comparing the quality of systems and software products.

Table 1. The Likert table for qualitative interpretation is used in interpreting the results of the survey.

| Scale | Range | Qualitative Interpretation |
|---|---|---|
| 1 | 1.00 – 1.80 | Poor |
| 2 | 1.81 – 2.60 | Fair |
| 3 | 2.61 – 3.40 | Good |
| 4 | 3.41 – 4.20 | Very Good |
| 5 | 4.21 – 5.00 | Excellent |

Table 1 is the Likert Scale used to rank or measure people's opinions on a subject matter or specific topic. It can also be used to measure the questions or result. The proponents used the Likert Scale to interpret the result and overall ISO/IEC 25010 evaluation average (Equation 1).

Formula:  ***Over-all Average = Total Average/Total Number of Items***   *Equation 1*

Table 2 is the ISO overall result for the application quality evaluation. By getting the total average, the researchers used the following formula.

Table 2. ISO Overall Result

| Software Product Quality Characteristics | Sub-Characteristics | |
|---|---|---|
| **Functional Suitability.** *This characteristic represents the degree to which a product or system provides functions that meet stated and implied needs when used under specified conditions.* | 1 | Functional Completeness |
| | 2 | Functional Correctness |
| | 3 | Functional Appropriateness |
| **Performance Efficiency.** *This characteristic represents the performance relative to the amount of resources used under stated conditions.* | 4 | Time-behavior |
| | 5 | Capacity |
| | 6 | Resource Utilization |
| **Compatibility.** *Degree to which a product, system or component can exchange information with other products, systems or components, and/or perform its required functions while sharing the same hardware or software environment.* | 7 | Co-existence |
| | 8 | Interoperability |
| **Usability.** *Degree to which a product or system can be used by specified users to achieve specified goals with effectiveness, efficiency and satisfaction in a specified context of use.* | 9 | Appropriateness Recognizability |
| | 10 | Learnability |
| | 11 | Operability |
| | 12 | User error protection |
| | 13 | User interface aesthetics |
| | 14 | Accessibility |
| **Reliability.** *Degree to which a system, product or component performs specified functions under specified conditions for a specified period of time* | 15 | Maturity |
| | 16 | Availability |
| | 17 | Fault tolerance |
| | 18 | Recoverability |
| **Security.** *Degree to which a product or system protects information and data so that persons or other products or systems have the degree of data access appropriate to their types and levels of authorization.* | 19 | Confidentiality |
| | 20 | Integrity |
| | 21 | Non-repudiation |
| | 22 | Accountability |
| | 23 | Authenticity |
| **Maintainability.** *Degree of effectiveness and efficiency with which a product or system can be modified to improve it, correct it or adapt it to changes in environment, and in requirements.* | 24 | Modularity |
| | 25 | Reusability |
| | 26 | Analyzability |
| | 27 | Modifiability |
| | 28 | Testability |
| **Portability.** *Degree of effectiveness and efficiency with which a system, product or component can be transferred from one hardware, software or other operational or usage environment to another* | 29 | Adaptability |
| | 30 | Install ability |
| | 31 | Replaceability |

# RESULTS

## System Architecture

The project's full system architecture diagram, shown in Figure 2 gives a visual representation of all of its components and how they relate to one another. In the case of the Road Updates Information System, this diagram depicts two (2) different user types: the first user is the System Admin, to whom all report approval is monitored and managed, and the second user is the user that represents the common people (such as riders and commuters, as well as citizens), to whom they can participate in reporting incidents in a particular road or location. Since the system leverages Google Maps to access existing maps with high accuracy, another entity in the diagram is the API, which is crucial. Last but not least, having connection to the Internet is crucial for using the system and maps because it is how the system connects to its database, accesses its user interfaces, and syncs in real time.

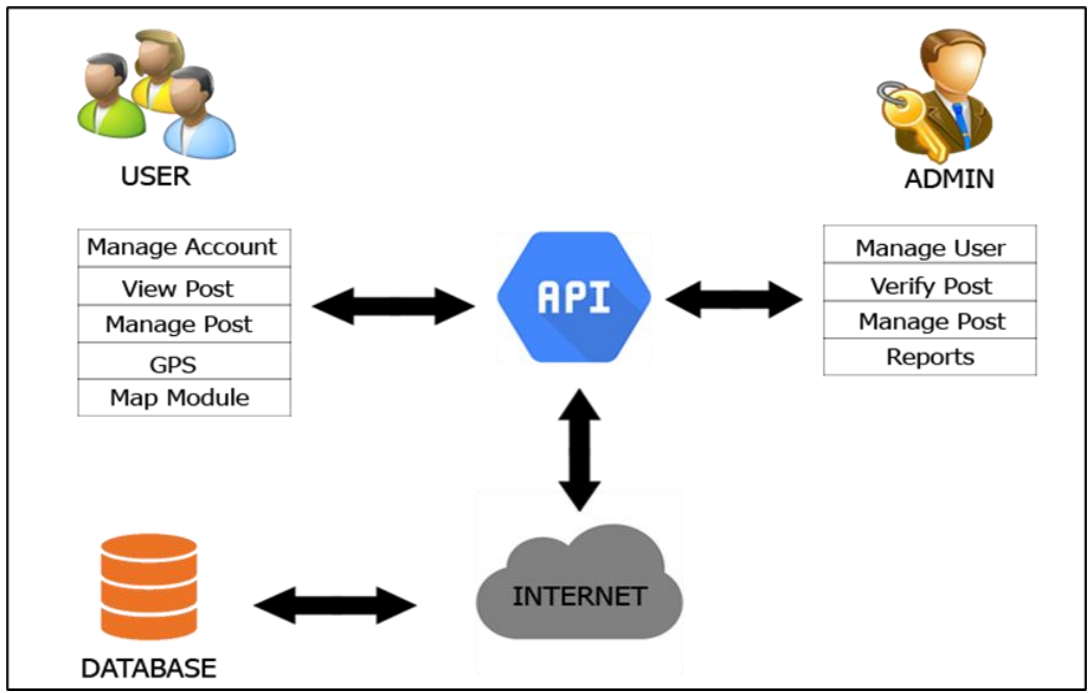

*Figure 2.* System Architecture Diagram

## System Login

A system login is utilized so that admin can access the system's primary dashboard (Figure 3). Only authorized system administrators were permitted to access the admin dashboard in order to validate report requests and maintain the system to protect the functioning and integrity of the data.

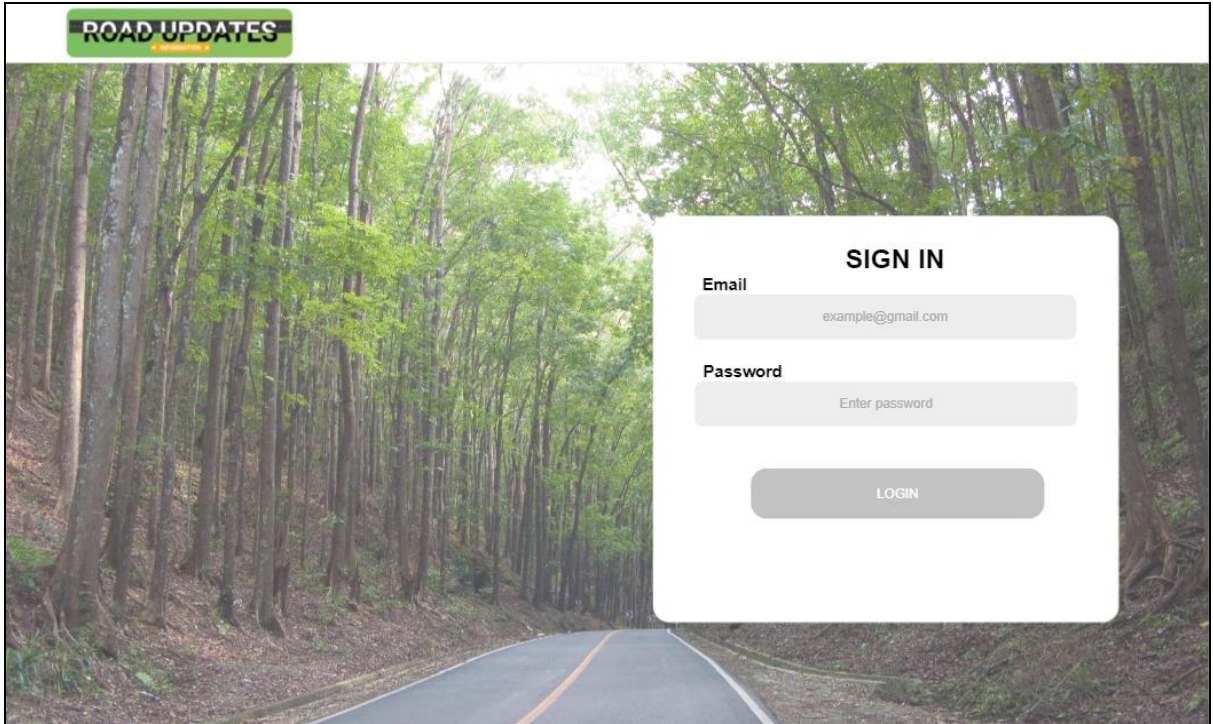

*Figure 3.* Admin Login

## Home Page

Figure 4 depicts the user's dashboard or home page, where road updates are displayed. This includes functionality for citizens who wish to report an incident by updating the map with information that is crucial, along with an attached photo to ensure that the actual incident is captured for visual presentation.

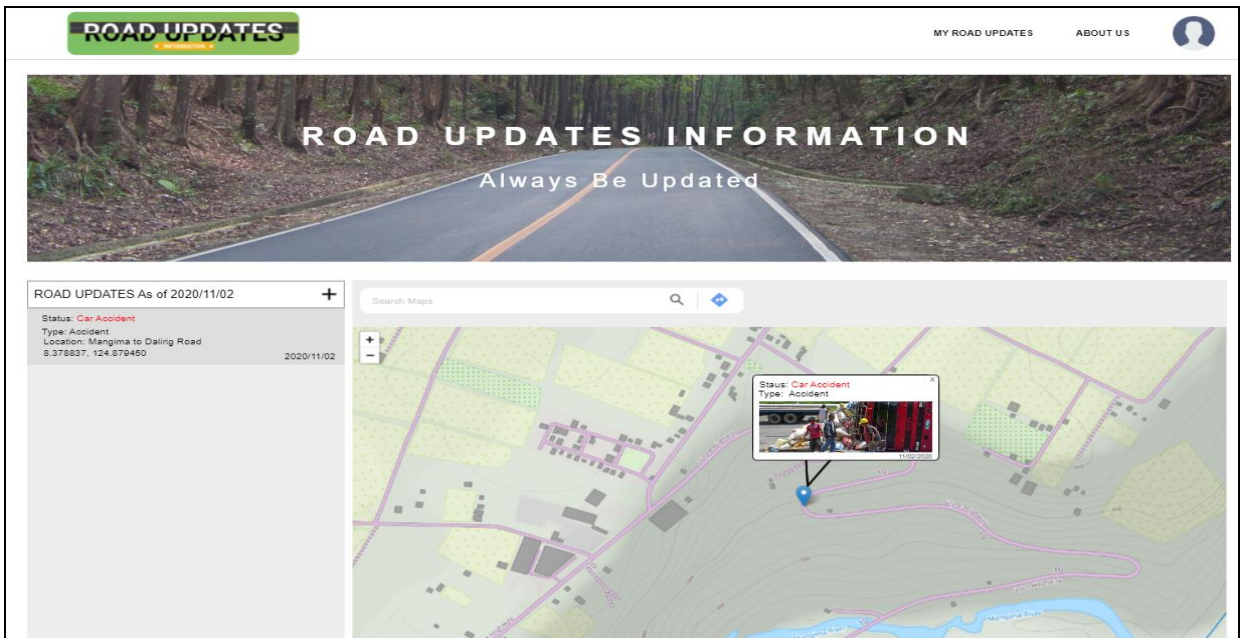

*Figure 4.* Homepage

## Report Information Portal

All users of the system have the option to submit road updates and other relevant incidents in order to notify others about the current and true status on a particular road utilizing this site, which is an example of "*Bayanihan,*" a Filipino character attribute. How to report a road update is shown in a modal in Figure 5. If possible, the individual reporting an incident should include attachments, preferably images of the actual event, that clearly describe the scenario as it actually is, the actual address, and the location point based on maps. The information from the road incident report will be sent to the admin report page for validation and approval.

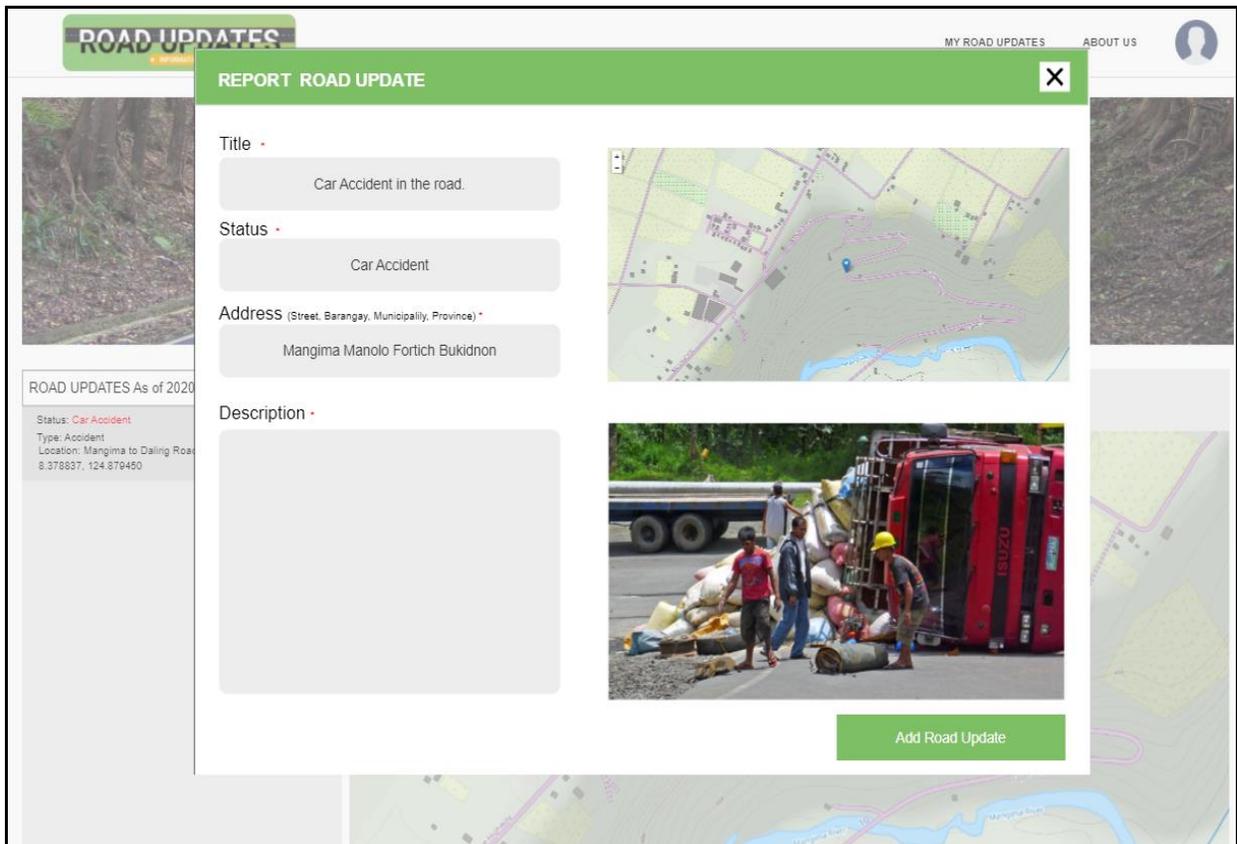

*Figure 5.* Report Information

## Admin Dashboard

Once a report has been completed and saved, the administrator will use an Admin Dashboard to monitor and verify it. The Admin Dashboard (Figure 6) shares much of the same content as the User Dashboard, with the exception that it has the ability to review reported road incidents and the discretionary power to approve when fully verified or to deny such reports when the reports are hoaxes, done solely to embarrass or prank people, or when the intention is uncertain.

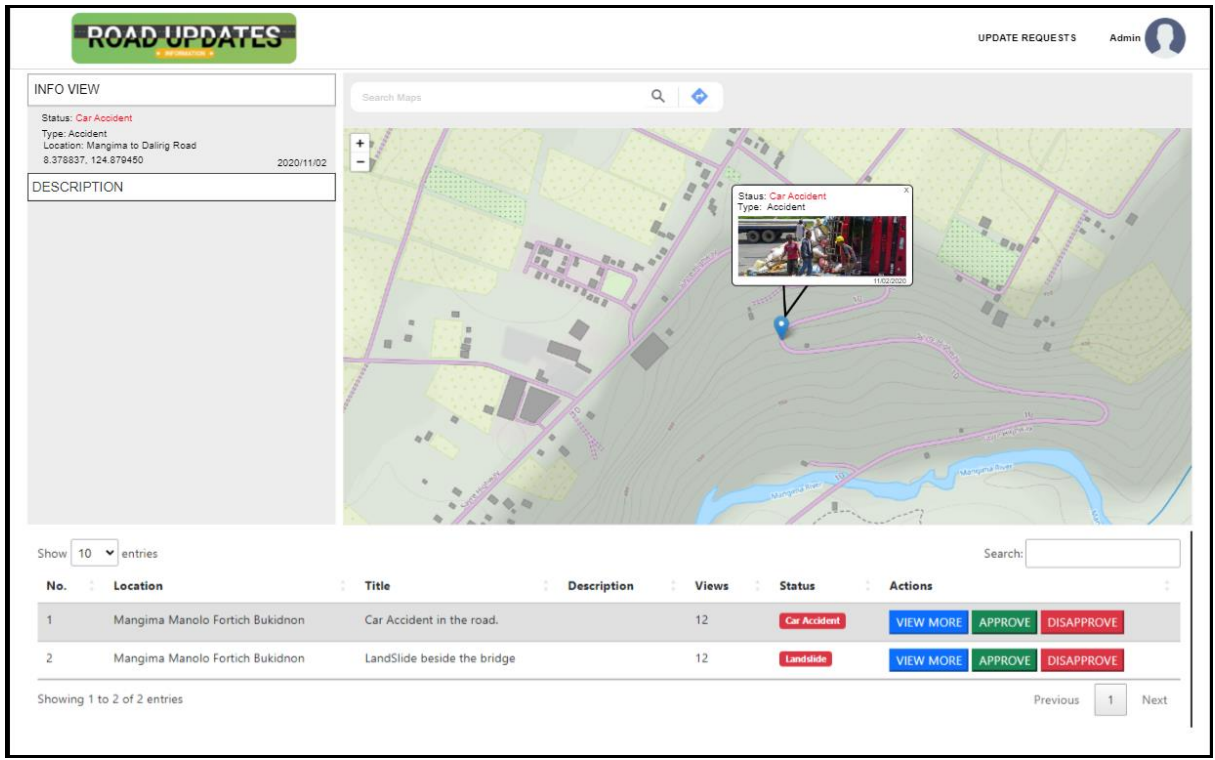

*Figure 6.* Admin Dashboard

## Evaluation Results

Table 3 displays the results in accordance with ISO/IEC 25010, which includes the quality characteristics that serve as the key criteria for assessing a system's or software's quality. The sub-characteristics that quantify the precise extent of each quality characteristic are included in the table. The Mean is the verbal interpretation of the mean of all user comments. To make the findings of this survey easy to recognize and comprehend, the replies were ranked using a likert scale. The system's success or failure will then be determined by the Overall Weighted Average based on the ISO standard, which is widely adopted globally. According to the table, the mean for each sub-characteristic is regarded as being between very good and excellent, and the overall weighted average is 4.21, which is interpreted as excellent on the likert scale.

Table 3. ISO/IEC 25010 Evaluation Result.

| Characteristics | Sub-Characteristics | Mean | Verbal Interpretation |
|---|---|---|---|
| Functional Suitability | Functional Completeness | 4.1 | Very Good |
| | Functional Correctness | 4.0 | Very Good |
| | Functional Appropriateness | 4.0 | Very Good |
| Performance Efficiency | Time-behavior | 4.0 | Very Good |
| | Capacity | 4.3 | Excellent |
| | Resource Utilization | 4.1 | Very Good |
| Compatibility | Co-existence | 4.0 | Very Good |
| | Interoperability | 3.8 | Very Good |
| Usability | Appropriateness Recognizability | 3.8 | Very Good |
| | Learnability | 4.6 | Excellent |
| | Operability | 3.8 | Very Good |
| | User error protection | 4.5 | Excellent |
| | User interface aesthetics | 4.0 | Very Good |
| | Accessibility | 3.8 | Very Good |
| Reliability | Maturity | 4.5 | Excellent |
| | Availability | 4.6 | Excellent |
| | Fault tolerance | 4.3 | Excellent |
| | Recoverability | 4.1 | Very Good |
| Security | Confidentiality | 4.1 | Very Good |
| | Integrity | 4.3 | Excellent |
| | Non-repudiation | 4.6 | Excellent |
| | Accountability | 4.8 | Excellent |
| | Authenticity | 4.0 | Very Good |
| Maintainability | Modularity | 4.0 | Very Good |
| | Reusability | 4.1 | Very Good |
| | Analyzability | 4.3 | Excellent |
| | Modifiability | 4.7 | Excellent |
| | Testability | 4.5 | Excellent |
| Portability | Adaptability | 4.5 | Excellent |
| | Install ability | 4.5 | Excellent |
| | Replace ability | 4.6 | Excellent |
| **Overall Weighted Average** | | **4.21** | **Excellent** |

# DISCUSSIONS

This study intent to design, develop and evaluate a web-based roads update information system as an instrument that can be used by anybody who wishes to know the current situation of a particular road in a specific place. Figure 2 shows the over-all System Architecture of the study, where we can see the conceptual model, which defines its structure, behavior, and more views of the developed system; in the case of this study, there are three (2) external entities that can access the API set by the proponents for front end use. For this system to function, a working internet connection is necessary in syncing the Front-end, API, and access to the data and information in the database.

System screenshots and system views are presented in Figures 3, 4, 5, and 6. Figure 3 requires the Login of an administrator to have access to the system to control malicious use and to verify that all contributors are legit users to avoid inaccurate updates. The Homepage in Figure 4 shows the home page of the web application by accessing through Smart Phone's browser, Laptop, and Computer. It is the landing page of a web application that the users can access. It contains the search location bar to search the old posts. Traffic alerts include the reports submitted to the users. The recent road updates include the type of incident, location, and time. Location maps are specialized maps dedicated to finding a specific place, while Figure 5 shows the report information page, allowing users to post information about a particular incident. The admin will review the report information to determine if the report is valid and concise. If the reviewed post were approved, it would be shown as a recent road update. Figure 6 shows the Admin Dashboard, where all approvals and management of the road updates are done. The Administrator has the discretion whether to approve or disapprove a particular report of road cases by a citizen because the admin needs to verify whether the claims are valid to avoid possible confusion if all case reports are just done without a reasonable basis.

In terms of the evaluation results, table 1 shows a very interesting result. The respondents of this study are drivers, commuters, and travelers. During the evaluation of the regular users, it gives the feedback score about the system of the final output. ISO/IEC 25010 is used for the application evaluation assigned to the participants. The developed web application was evaluated and highly accepted average which is 4.21. Functionality suitability with a mean of 4.00, which is "Very Good", the degree to which a product or system provides functions accomplished defined and suggested needs when used under specified conditions; performance efficiency has a mean of 4.0, which is "Very Good", the performance relative to the amount of resources used under stated conditions; compatibility has a mean of 4.0, which is "Very good" the system can interchange details with other goods, systems/structure or elements, and/or carryout its necessary functions while sharing the same hardware or software environment, usability has a mean of 4.6 which is "Excellent", the system can be used by specified users to achieve established goals with efficacy, efficiency and satisfaction in a specified context of use, reliability has a mean of 4.5 which is "Excellent", It performs designated functions under specified state for a specified period of time, security has a mean of 4.5 which is "Excellent", the system safeguard information and data so that entities such as persons or other commodity or systems/modules have the standard of data access suitable to their types and levels of authorization. , maintainability has a mean of 4.5, the potency and efficiency with which a product/commodity or systems/modules can be change to improve, correct, or adapt it to changes in the environment, and requirements and portability have a mean of 4.5 which is "Excellent"- the system can be transferred from one hardware, software or other operational or usage environment to another. The overall average weighted mean of the application is 4.21, which was rated as "Excellent". This means that application is acceptable during the evaluation.

## CONCLUSIONS

The developed web application meets the objectives to provide the users. The web application is well-recommended and acceptable by the users. The researchers provided cheaper premium web hosting every year for its renewal. The researchers will incorporate ads for additional revenue for the web application and a registration form for the users to get an acknowledgment of their contribution to the post and to improve the web application. Agencies such as DPWH, RTA, and the like should be able to contribute accurate reports using the web application to ensure reliable and well-grounded information is delivered to the community.

## RECOMMENDATIONS

It is recommended that the future development expand the coverage of the road updates, including the entire Philippine archipelago, for long-drive commuters and drivers to be more updated regarding road updates. Also, include using mobile applications for more user-friendly design and interactions.

## RESEARCH IMPLICATIONS

The study allows viewing and reporting road updates with attachments as proof, informing commuters about what to expect when they travel. The system has been implemented in the municipality of Bukidnon and has been updated now and then for continuous quality improvement.

## PRACTICAL IMPLICATIONS

The main benefit that the Road Updates Information System can do is to inform people who are constantly traveling to be updated on the current situation of a specific road by accessing the system and checking the map on areas to travel and see for themselves the photos and description of a place where road construction, accidents and other causes that delay the traveling time of passengers and travelers. Aside from being informed of the road updates, citizens can also participate by uploading photos and descriptions of what is still going on the roads to the system and be viewed by people who try to access the web portal.

# DECLARATIONS

## *Conflict of Interest*
No author has disclosed any conflicts of interest.

## *Informed Consent*
The study did not involve humans as participants and only used scene text datasets that are already available online hence this is not applicable.

## *Ethics Approval*
The conducted research uses only scene text datasets, which are already available online, and did not include humans as participants; hence this is not applicable.

## Authors' Biography

**Benzar Glen S. Grepon** is the Director of the Information Management Office and the previous IT Program Head of Northern Bukidnon State College, specializing in research, Programming, Project Management, Management Information Systems, Network Management, and Administration.

**JC P. Margallo** is a Bachelor of Science in Information Technology graduate College at Northern Bukidnon State College, specializing in Web development and programming.

**Jonathan B. Maserin** is a Bachelor of Science in Information Technology graduate College at Northern Bukidnon State College, specializing in Web development and programming.

**Rio Al-Di A. Dompol** is a Bachelor of Science in Information Technology graduate College at Northern Bukidnon State College and a computer programmer in the Information Management Office of NBSC, specializing in Web development, programming, Database Management and Mobile Design and Development.